\documentclass[twocolumn,eqsecnum,aps]{revtex4}
\usepackage{graphicx}

\newcommand{\be}{\begin{equation}}
\newcommand{\eq}{\end{equation}}
   
\begin{document}

%
\title{$SU(N_c \to \infty)$ Lattice Data and Degrees of Freedom
of the QCD string.}
\author{S. Dalley}
\affiliation{Department of Physics, University of Wales Swansea, 
Singleton Park,
Swansea SA2 8PP, United Kingdom}
\preprint{SWAT/05/453}
\pacs{}

\begin{abstract}
Lattice simulation data on the critical temperature
and long-distance potential,  that probe the 
degrees of freedom of the QCD string, are critically reviewed. 
It is emphasized that comparison of experimental or $SU(N_c)$ lattice 
data, at finite number of colors $N_c$, with free string theory can be 
misleading due to string interactions. Large-$N_c$ 
extrapolation of pure lattice gauge
theory data, in both 3 and 4 dimensions,
indicates that there are more worldsheet degrees of freedom than the purely 
massless transverse ones of the free
Nambu-Goto string. The extra variables are consistent
with massive modes of oscillation that effectively contribute like 
$c \approx 1/2$
conformal degrees of freedom to highly excited states.
As a concrete example, the highly excited spectrum of the 
Chodos-Thorn relativistic string in $1+1$ dimensions is analyzed, 
where there are no
transverse oscillations.
We find that the asymptotic density of states for this 
model is characteristic of a $c=1/2$ conformal worldsheet theory.
The observations made here should also constrain the backgrounds of
holographic string models for QCD.

\end{abstract}

\maketitle



\section{Introduction}
\label{intro}

It is well-known that the long-wavelength  properties of confining gauge
theories can be 
modelled quite accurately by thin, structureless relativistic strings
oscillating transverse to themselves. The prototypical example is the 
Nambu-Goto model \cite{nambu}, with action proportional to worldsheet area.
 At shorter wavelengths, 
additional properties or degrees of freedom are likely to play a role. 
A related
and popular 
suggestion has been that the string is always thin and structureless, 
but that the additional effects can effectively be seen  
as motion in a curved  higher dimension 
\cite{maldacena,polyakov}.
One would like to characterize any deviations from structureless
strings in spacetime in order to determine what, if any, string theory exactly
describes confining gauge theories at all length scales. 
In this paper, data from lattice simulations of $SU(N_c)$
pure gauge theories will be used as a guide. 
It is only quite recently that such data have become accurate and
extensive enough to be useful in this regard. Based on this analysis,
a concrete 
suggestion is made for additional degrees of freedom that must be included
at short wavelengths --- 
longitudinal modes of oscillation similar
to those of the Chodos-Thorn massive string \cite{chodos}.

The gauge theory data used are the asymptotic 
density of states in the
hadron spectrum, the (related) Hagedorn temperature, and the 
long-distance groundstate potential energy 
of a winding string or pair of heavy sources. These probe the short and
long wavelength regimes respectively. The existence in gauge theory
of a Hagedorn temperature in particular --- 
which is to be distinguished from the 
deconfinement temperature in general ---  is an important indication that a 
string description exists on all length scales.
It is not the intention to provide here a 
complete historical review of lattice data;
only the most accurate and relevant results have been selected.
Data from 3 as well as 4
spacetime dimensions are used, since pure gauge theory seems to be described by
a string theory in both cases.
Data for fixed, finite $N_c$, which are often used to 
address such questions, can be misleading if compared with a free string
model. It is shown that string interactions of strength $1/N_c$ can mask 
the effects of
additional worldsheet degrees of freedom. To identify the latter
unambiguously, one must extrapolate data to $N_c=\infty$.
Large-$N_c$ results are shown to 
indicate a Hagedorn temperature below that of the
Nambu-Goto free string, equivalent to adding an extra effective
conformal degree
of freedom with central charge $c \approx 1/2$. However, the 
``Luscher-term'' \cite{luscheretal,luscher},
in the asymptotic $1/l$ expansion of the groundstate
energy of a gauge string with minimal
allowed length $l$, seems to be consistent with the Nambu-Goto result
and not to depend on $N_c$. These two statements are consistent if
the additional worldsheet degrees of freedom, beyond those of Nambu-Goto,
are massive. Only at energies large compared to their mass is a 
conformal approximation valid. 

In the second part  of the paper, it is suggested that massive longitudinal 
oscillations are responsible for the extra
degrees of freedom. This is hardly an original observation. However, 
new evidence is presented that such oscillations have roughly the right
number of degrees of freedom to account for the facts displayed in the first
part of the paper. The Chodos-Thorn massive relativistic string
is essentially the Nambu-Goto model with additional massive pointlike
insertions of energy-momentum on the string. Although this
model is difficult to solve in greater than two dimensions, 
the asymptotic spectrum is found exactly in two dimensions, where only 
longitudinal oscillations remain.
It is found that these are equivalent to those
 of a $c=1/2$ worldsheet field theory. 
Whether this is coincidence and we are barking up the wrong tree
remains to be seen. In any event, the lattice data  strongly constrain any
purported string theory of QCD.

\section{$SU(N_c \to \infty)$ data}

\subsection{Stringy observables}

Let us first briefly review the physical observables used later. 
The boundstate spectra of string  theories 
generically have a density of states,
$\rho(M)$ per unit interval of mass $M$, that grows exponentially
for large $M$.
If, on the worldsheet of the {\em free} string, there is a conformal field
theory of central charge $c$ representing physical oscillations (not 
counting those elimated by reparameterization invariance etc.), then 
as $M \to \infty$ \cite{vafa}
\be
\rho(M) \propto M^{-(3+D_\perp)/2}
{\rm exp} \left(\frac{M}{T_H}\right) \ ,
\label{exp}
\eq
where 
\be
T_H = \sqrt{\frac{3 \sigma}{c \pi}} \ ,
\eq
$\sigma$ is the string tension, and $D_\perp$ is the effective number of
dimensions for transverse oscillations.
Note that the same asymptotic density of states can be expected even
if the worldsheet theory is massive, once the excitation energies
are far above the relevant mass scale. The theory is effectively
conformal in this regime.
As is well-known, the canonical partition function of the free string
gas diverges above the Hagedorn temperature $T > T_H$ \cite{hagedorn}. 
The physics of this point
is a second order phase transition \cite{parisi} 
--- we assume the power law corrections in 
eq.~(\ref{exp}) are such that the internal energy does not diverge at $T_H$ ---
driven by the entropy of strings.
The transition is also 
signalled by the vanishing of the mass $E_c(1/T)$ of the  string
that winds once around the compact Euclidean time direction of circumference
$1/T$ in the 
finite-temperature partition function \cite{olesen}. 
All these statements have an analogue in confining gauge theories.
The (real) hadronic spectrum does rise exponentially \cite{broniowski}
and there are phase
transitions in pure gauge theory
at which winding Polyakov loops get a VEV \cite{polyakov2}. 
When those transitions
are second order, the Polyakov loop mass vanishes and one may assume
this is a  Hagedorn transition at $T=T_H$. If the gluon entropy is more 
important than the string entropy, which tends to be the case for larger
$N_c$ \cite{1storder,bern,teper05}, 
a first order transition at $T_c<T_H$ will occur first.
However, it has been shown by Teper and Bringoltz \cite{bringoltz}
that one may study the approach to $T_H$ in the metastable superheated phase
above $T_c$;
in fact, this phase should become stable when $N_c \to \infty$. 

Another measure of the degrees of freedom of string theory results
from the asymptotic $1/l$ expansion \cite{luscheretal} of the groundstate
energy $E_o(l)$ of an open string with
endpoints at fixed separation $l$:
\be 
E_o(l) = \sigma L + {\rm const.} -\frac{\pi\tilde{c}}{24 l} + \cdots \ .
\label{open}
\eq
The coefficient of the ``Luscher-term'',
with  $\tilde{c} = c - 24h$, is related
to the Casimir energy of free massless fields \cite{luscher}
of central charge $c$
and the lowest dimension $h$ of primary field that propagates on the
worldsheet. There is a similar expression for a closed
string \cite{deF} that winds once around a  compact  spatial direction of
circumference $l$:
\be 
E_c(l) = \sigma L + {\rm const.} -\frac{\pi\tilde{c}}{6 l} + \cdots \ .
\label{closed}
\eq 
The analogues of these configurations in
pure gauge theory are the potential between heavy sources in the fundamental
represenation of $SU(N_c)$ and the Polyakov loop mass. As well as the
$l \to \infty$ limit, we will also be interested in the minimum allowed
$l$, since $E_c(1/T_H) = 0$ can be taken to define the Hagedorn temperature.

\subsection{Why $3 \neq \infty$}

The original connection between string worldsheets and confining gauge
theories was made by 't Hooft \cite{hoof}, 
in the case of weak gauge coupling $g$
expansion, and by Wilson \cite{wilson}, 
in the case of strong gauge coupling expansion.
In both cases, the string coupling $g_s$, by which we mean the (logarithm 
of the) coupling to the worldsheet topological invariant  
\be
\int dx^0 dx^1 \ R \sqrt{- G}  \ , 
\label{top}
\eq
is $1/N_c$. Here,  $G= \det G_{\alpha \beta}$, where the 
induced worldsheet  metric is 
\be
G_{\alpha \beta} = \frac{\partial X^\mu}{\partial x^\alpha}
\frac{\partial X_\mu}{\partial x^\beta} \ ,
\eq
$R$ its Gaussian curvature, and $X^\mu$ the 
worldsheet embedding co-ordinate in spacetime of dimension $D$, 
$\mu \in \{0,\cdots D-1\}$, and $\{x^0,x^1\}$ the intrinsic
coordinates on the worldsheet.   
Eq.~(\ref{top}) governs the splitting and joining interactions of strings. 
In the modern
era also, the ADS/CFT correspondence \cite{maldacena}
and its non-conformal non-supersymmetric
generalisations 
are between $SU(N_c)$ gauge theories and
fundamental string theory with coupling $g_s \propto 1/N_c$ in the $N_c \to
\infty$ limit.
If we are trying to phenomenologically 
determine the rest of the worldsheet action that
must be added to (\ref{top}) in gauge theories, 
the presence of string interactions greatly
complicates matters. Most results for string theories 
have been derived to leading orders in the string coupling expansion.
In particular, the spectrum of string states is known exactly for several
string actions, but only for free strings. 

The Nambu-Goto model adds
to the worldsheet action the area term:
\be
- \sigma \int dx^0 dx^1 \sqrt{-G}.
\label{NG}
\eq
By lightcone gauge fixing of reparameterization invariance and 
Fourier decomposition of transverse coordinates ($i \in \{ 1 , \cdots D-2\}$)
for open strings
\be
X^i = X^{i}_0 + P^i x^0 + {\rm i} \sum_{n \neq 0}
 \alpha_{n}^{i} {\rm e}^{-{\rm i} n x^0} \cos nx^1 \ ,
\eq
free strings ($g_s = 0$)
in the quantized model have mass spectrum operator
given by a sum of harmonic oscillators \cite{thorn} 
\be
M^2 = 2 \pi \sigma \sum_{n=1}^{\infty} 
\alpha_{-n}^{i} \alpha_{n}^{i} \ , 
\label{spec}
\eq
\be
[\alpha_{n}^{i}, \alpha_{m}^{j}] = n \delta_{nm} \delta^{ij} \ .
\eq
The Nambu-Goto model is a classical conformal field theory in two dimensions
with $c= D-2 = D_\perp$, $h=0$. Therefore, it predicts 
\be
\tilde{c} = D-2 \ ,\ T_H = \sqrt{\frac{3 \sigma}{ (D-2) \pi}} \ . 
\eq 
In particular
\begin{eqnarray}
T_H & \approx & 0.98 \sqrt{\sigma} \ (D=3) \ ,\label{d3}\\
T_H & \approx & 0.69 \sqrt{\sigma} \ (D=4) \ .\label{d4}
\end{eqnarray}
Although this model is not Lorentz invariant for $D \neq 26$, the action
should probably be understood as the long distance approximation
of a more general one
describing gauge theory strings.
In principle one can include further dimensionful interactions
between the transverse degrees of freedom $X^i$ 
or add further degrees of freedom. 
This procedure can be made formal through the $1/l$
expansion \cite{luscheretal}, and there is some indication that
consistency problems can be fixed order-by-order \cite{polchinski}. 
If the additional degrees of freedom 
are massive, they could be integrated out systematically
to leave higher order corrections to the Nambu-Goto action.

At first sight, it appears that the simple Nambu-Goto
model, in the case of free strings, is in excellent agreement with
data on the observables mentioned above in $SU(3)$ gauge theory (the case
relevant for QCD). Firstly, recent precision lattice data on the value
of the Luscher coefficient,
indicate that $\tilde{c} = D-2$ \cite{luscher2,sommer}. Indeed, 
this also appears to be the case
for other gauge groups, including
 $Z_2$ \cite{caselle,wako}, $SU(2)$ \cite{lucini,wako}, and 
$SU(N_c \to \infty)$ \cite{meyer}, suggesting
the $1/N_c$ string interactions do not affect the simple Casimir effect
argument for the coefficient of the Luscher term. 
In principle, it is possible that there are additional massless degrees 
of freedom,
$c > D-2$, but still $\tilde{c} = D-2$. Although this cannot technically
be ruled out, it seems an unlikely coincidence and, moreover,
the groundstate is usually not a scalar if $h>0$.
More generally,
the result would  be consistent with additional dimensionful 
worldsheet interactions or additional massive degrees of freedom, since 
these should not contribute to the long-distance $l$ properties of the string.

Pure lattice gauge theory simulations of the Polyakov loop correlators
enable accurate determinations of the transition temperature, with 
the loop expectation value as order parameter.
For $D=3$, $SU(3)$ pure lattice gauge theory the transition
is 2nd order and  agrees with the $D=3$ Hagedorn temperature of the 
simple Nambu-Goto model (\ref{d3}) \cite{lattice,teper93}. 
For $D=4$ and $SU(3)$, the deconfinement transition temperature $T_c$ 
is slightly below the Nambu-Goto
Hagedorn temperature (\ref{d4}) \cite{1storder}. 
But it is weakly first
order, so the actual gauge theory Hagedorn temperature is probably slightly 
higher than $T_c$. 
Although it is difficult to study the density
of hadron states directly in lattice gauge theory, except in some effective
theories \cite{prl}, 
this data can be extracted from experiment. The meson resonance
spectrum shows a clear exponential rise.
A study by
Dienes and  Cudell \cite{dienes}
establised that the best fit to the data was for 
$c= D_{\perp}$ = 2, in agreement with the Nambu-Goto model. 

\begin{figure*}
$\displaystyle T_c/\sqrt{\sigma}$\hspace{5pt}
\raisebox{-1.75in}{\includegraphics[width=4in]{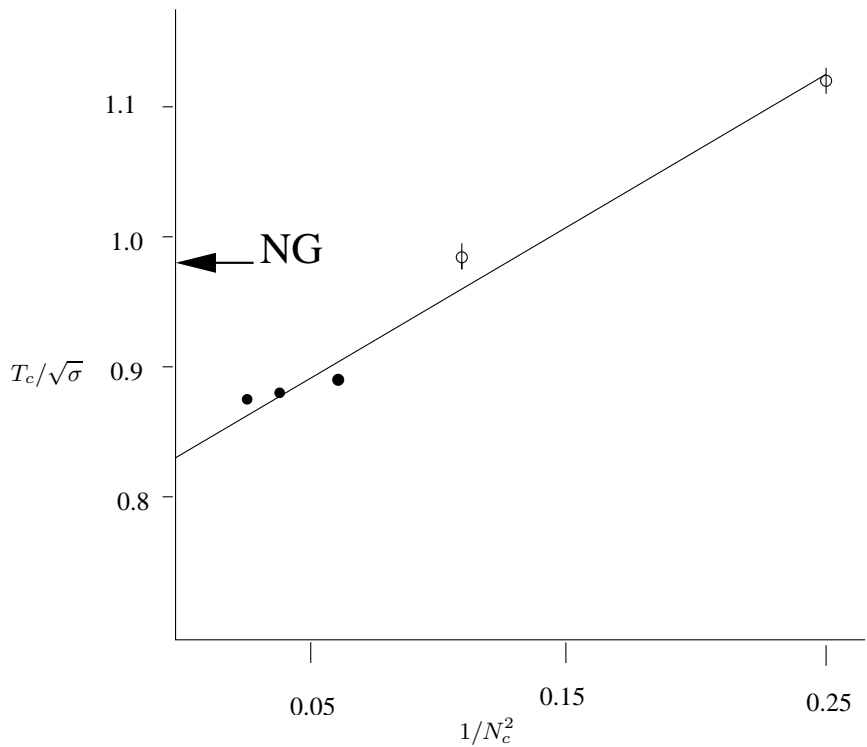}}\\
\hspace{0.5in}$\displaystyle 1/N_{c}^{2}$
\caption{The variation of the transition temperature $T_c$, in 
units of the zero-temperature
string tension $\sigma$, with the number of colours $N_c$ for pure
$SU(N_c)$ gauge theory in $D=3$ dimensions. Open circles \cite{teper93} 
and
filled circles \cite{liddle} are from different simulations. 
The linear fit is to the 2nd
order transitions at $N_c = 2,3, 4$, where we identify $T_c = T_H$. NG 
indicates the free-string Nambu-Goto prediction.  
\label{fig1}}
\end{figure*}

\begin{figure*}
$\displaystyle T_c/\sqrt{\sigma}$\hspace{5pt}
\raisebox{-1.75in}{\includegraphics[width=4in]{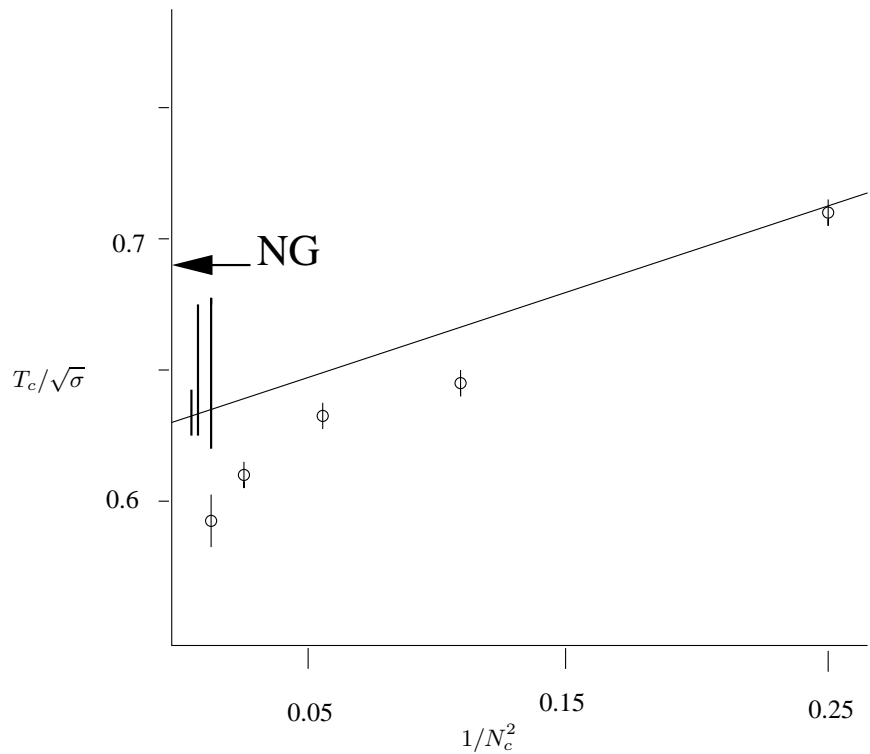}}\\
\hspace{0.5in}$\displaystyle 1/N_{c}^{2}$
\caption{Open circles show the variation of the deconfinement
temperature $T_c$, in units of the zero-temperature
string tension $\sigma$, with the number of colours $N_c$ for pure
$SU(N_c)$ gauge theory in $D=4$ dimensions \cite{teper05,1storder}. 
The solid bars represent the range of Hagedorn temperatures obtained
in ref.\cite{bringoltz} by different fits to the Polyakov loop mass.
The straight line fit is to these and the 2nd order $SU(2)$ 
transition, where we identify $T_c = T_H$. 
NG  indicates the free-string Nambu-Goto prediction.  
\label{fig2}}
\end{figure*}

\begin{figure*}
$\displaystyle \ln t$\hspace{5pt}
\raisebox{-1.75in}{\includegraphics[width=4in]{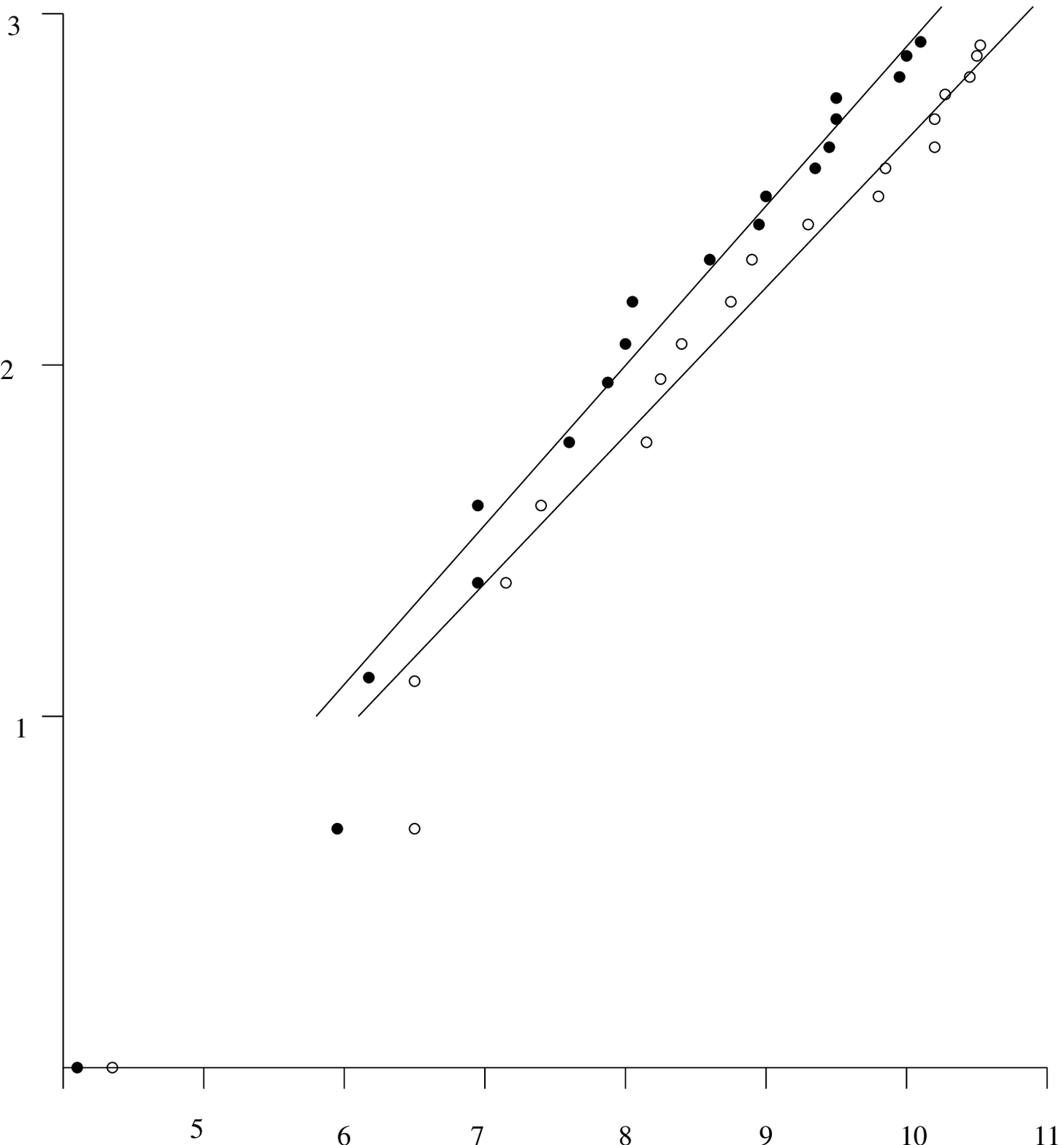}}\\
\hspace{0.5in}$\displaystyle M_t/\sqrt{\sigma}$
\caption{The mass $M_t$ of the $t^{\rm th}$ glueball in the spectrum 
of pure $SU(2)$ (open circles) and $SU(\infty)$ (filled circles) gauge
theory in $D=3$ dimensions \cite{teper98}. 
The lowest 19 states are shown. The straight
lines are to guide the eye only.
\label{fig3}}
\end{figure*}

Given this evidence, one might be led to conclude that the string theory
relevant for QCD is just the Nambu-Goto one. Of course, this cannot be true
because the model is not even consistent in 3 and 4 dimensions. The spectrum
of the $SU(3)$ heavy-source potential shows clear 
deviations from the Nambu-Goto
prediction for small separations \cite{morning}. 
Therefore it would be suprising if the Hagedorn temperature were correctly
predicted.
Results for other gauge groups, however, imply a more 
complicated picture. For $D=3$ pure gauge theory,
the transition  is second order for $N_c = 2,3,4$ (or at worst 
very weakly first order for $SU(4)$, so the Hagedorn point is 
very close by), and first order for higher $N_c$ \cite{teper93,liddle,holland}.
The corresponding temperatures are plotted versus $1/N_c$ in Figure~1.  
If we are correct in identifying the $N_c = 2,3,4$ points as Hagedorn
transitions, there is clearly a strong dependence of $T_H$ on $N_c$. 
For $D=4$ pure gauge theory, the transition is 2nd order only for $SU(2)$
\cite{teper05}.
However, Teper and Bringoltz \cite{bringoltz}
have recently been able to follow the Polyakov
loop mass into the metastable phase above the 1st order transitions
for $N_c > 2$, identifying the Hagedorn temperature
from $E_c(1/T_H) = 0$.  Figure~2 shows again that $T_H$ depends upon
$N_c$. 
Recalling that this temperature in a free resonance gas 
is dictated by the asymptotic density of states, it is natural to conclude
that the $1/N_c$ string interactions alter the spectrum so as to reduce the
density of states; they will also give states a width, but this is not usually
noticed in a lattice simulation. In fact,
this thinning of the spectrum 
can be seen already in the low-lying glueball spectrum.
Figure~3 compares the lowest glueball masses for $SU(2)$ 
and $SU(N_c \to \infty)$ 
in 3 dimensions \cite{teper98}. For $\rho(M)$ to decrease, the mass shift 
of the ${\rm t}^{th}$ glueball should increase with $M_t$.
A similar qualitative 
effect can be discerned in the spectrum 
for $D=4$ \cite{meyer2}, although fewer accurate
glueball masses are available in this case.

The results for the $SU(3)$ Hagedorn temperature
thus appear to be accidentally close to those for the
free Nambu-Goto string. The strings of an $SU(3)$ gauge theory are not free,
but have $1/N_c$ interactions which mask worldsheet effects
not accounted for by the Nambu Goto model. 
To compare like-with-like, one should 
only use free string formulas when comparing with the $N_c \to \infty$
limit  of gauge theory. What do such extrapolations predict? In pure
gauge theory, one expects physical observables to be expandable
as an aymptotic series in $1/N_{c}^2$ \cite{hoof}.
If one naively extrapolates the Hagedorn temperatures
of pure gauge theory to $N_c = \infty$, using fits $A + B/N_{c}^2$, one finds
\begin{eqnarray}
T_H & \approx & 0.83 \sqrt{\sigma} \ (D=3) \label{3D} \\ 
T_H & \approx & 0.63 \sqrt{\sigma} \ (D=4) \label{4D} \ . 
\end{eqnarray}
Note that, unlike the papers from which the data is sourced,
the fit is only to the second-order transitions. 
Since they arise from a separate
physical mechanism,  there is no
reason for the  first order points to be analytically related to them. 
In general,
there is also no theoretical reason why one should truncate the fit at a simple
$1/N_{c}^{2}$ correction; this is dictated only by the paucity of data.
Given that the fit is taken down to $N_c=2$, corrections at $O(1/N_{c}^{4})$ 
are likely to be the largest source of
error in the numbers (\ref{3D})(\ref{4D}).  
Comparing these with eq(\ref{exp}), in both cases this predicts an effective
$c \approx (D-2) + \frac{1}{2}$ worldsheet theory contributing to 
highly excited states. Given that the Luscher coefficient seems
unaffected by string interactions and agrees with the Nambu-Goto
result, whatever degrees of freedom give rise to the ``$+\frac{1}{2}$''
must be massive, although effectively conformal at high energies.

Given the volatility of numerical lattice results, this neat picture could
be upset, of course. There is already some inconsistency in the $D=3$
results for transition temperatures, since the extrapolation of 
2nd order transitions at $N_c = 2,3,4$ seems to lie below the first 
order transitions at $N_c = 5,6$ (also shown on fig~1), which cannot be right.
A study of the superheated phase at higher $N_c$ would be useful
in this respect. The first order transitions
for $D=4$ and $N_c > 2$, on the other hand, lie correctly below the
``Hagedorn line'' in fig~2.
While the ``$+\frac{1}{2}$'' may vary in size in the light of further 
simulations, it is clear that present data are accurate enough
to demonstrate further string degrees of freedom with approximately
this contribution.

\section{Longitudinal string oscillations}

\subsection{Spectrum}

In this section, we present evidence that massive longitudinal string
oscillations may account for the shift in the effective
central charge in the highly excited spectrum. Longitudinal 
degrees of freedom
are obviously not a new idea. In a sense, the Nambu-Goto model already
posseses them if $D \neq 26$. Both the covariant quantization \cite{goddard}
and the original Polyakov formulation (Liouville theory) 
\cite{polyakov3} lead to an extra $c=1$ (massless)
degree of
freedom corresponding to longitudinal oscillation. This is too much; it will
add to the Luscher coefficient; it will produce too many degrees of
freedom for the asymptotic spectrum (in the Polyakov case, the Liouville
zero mode produces a continuous spectrum).
But massive longitudinal modes may be expected. For example, 
they occur naturally in the
Nielsen-Olesen vortex solution of the Abelian Higgs model \cite{nielsen}. 
Also the Polchinski-Strominger effective string action indicates that
the Liouville field gets a mass \cite{polchinski}.
Here, 
a specific model with longitudinal oscillations wiil be analysed --- the 
excited spectrum of the Chodos-Thorn \cite{chodos} 
massive relativistic string. 
Unfortunately, this model is difficult to solve above 2 dimensions. But the
longitudinal modes of interest in are present in 2 dimensions
and the asymptotic spectrum can be found exactly in this case. 
Assuming that coupling
of longitudinal and transverse degrees of freedom in higher dimensions
does not drastically alter things, this will provide a measure of the
number of longitudinal degrees of freedom. 

\begin{figure*}
\raisebox{-1.75in}{\includegraphics[width=4in]{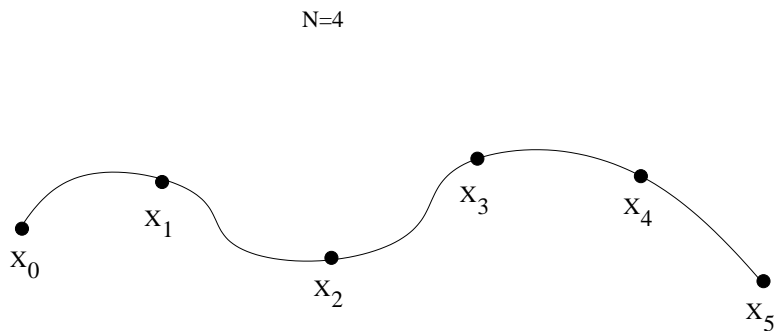}}\\
\caption{Open string with particle insertions ($N=4$ in this example).
\label{fig4}}
\end{figure*}

The action generalises the Nambu-Goto one (\ref{NG}) by the insertion
of massive particle degrees of freedom on the string
\be 
-  \int dx^0 dx^1 \left( \mu \sqrt{\partial_{0} X^\mu 
\partial_{0} X_\mu}+ 
\sigma \sqrt{-G} \right) \ .
\eq
In principle, the mass distribution could be continuous along the string,
but we will consider the case when it is discrete. Such a string is likely
to have the power-law fall-off of high energy scattering amplitudes
characteristic of particle field theory, which does not occur with conventional
strings in general \cite{gross}.
In addition to transverse oscillations, there are now oscillations 
associated with motion of the point masses along the length of the 
string. 
A semi-classical analysis of the free string spectrum of this
theory in $D=2$ dimensions 
was performed by Bardeen {\em et al.} \cite{bardeen}. 
In the sector with pointlike  
insertions at positions $X_j$, $j \in \{1, \cdots, N\}$,
in the interior of an open string, and insertions at each end
$X_0$, $X_{N+1}$ (see Figure~4), they found the  lightcone gauge Hamiltonian
\be
P^- = \sum_{j=0}^{N+1} \frac{\mu^2}{2 P^{+}_{j}}  + \sum_{j=0}^{N}
\sigma |X^{-}_{j} - X^{-}_{j+1}| \ . 
\label{ham}
\eq
In two dimensions, the entire lightcone momentum of the string is carried by
the insertions
\be
P^+ = \sum_{j=0}^{N+1} P^{+}_{j} \ .
\eq 
Finding the 
normal modes of classical solutions in the $\mu \to 0$ limit, Bardeen {\em et 
al.}
then impose Bohr-Sommerfeld
quantization conditions to obtain a spectrum of masses  
\be
 M^2 = 2 \pi \sigma \sum_{n=1}^{N+1} n (l_n + {\rm const.}) \ ,
\label{sc}
\eq
where $l_n \in \{0,1,2, \cdots \}$. 
Note that the massless limit $\mu \to 0$ is {\em not} the 
Nambu-Goto model, provided we allow points on the string, where there were 
masses, to move at the 
speed of light. Although the actions are the same, the Hamiltonians are
not \cite{bardeen}.

An important question is whether one should include sectors of different
$N$ independently in the spectrum (\ref{sc}).
In the $\mu \to 0$ limit, the classical solutions of the equations 
of motion  for an open string with $N$ insertions contain
the solutions for $N' < N$ insertions also. This happens because a subset
of the normal-mode motions occur with some of the insertions at coincident
points. For example, one of the $N=1$ classical solutions has the `interior'
point always attached to one or other end of the open string, 
$X_1 = X_0$ or $X_1 = X_2$. Such a motion is geometrically indistinguishable
from the classical solution for $N=0$. Therefore, one might expect that
in this case the entire spectrum can be obtained by taking $N \to \infty$
in eq.(\ref{sc}).
If one compares this spectrum with eq.(\ref{spec}) and recall that 
there are no transverse oscillations for $D=2$, the longitudinal 
oscillations in this model as $N \to \infty$ appear to 
asymptotically give
the same spectrum as a physical $c=1$ worldsheet degree of freedom 
(like one extra transverse dimension). However,
once a mass $\mu$
is introduced, however small, the indistinguishability is lost; the
total mass of the insertion is significant. For this reason, the sectors
for each, fixed $N$
give rise to distinguishable states in the spectrum, resulting in many
more states than just the $N \to \infty$ sector. In fact, if all $N$ sectors
are allowed, the presence
of modes with $l_n =0$ would lead to each energy level in the spectrum
being infinitely degenerate in the $\mu \to 0$ limit
(neglecting the constant in (\ref{sc})). 

The resolution of this difficulty lies in the fact that
one expects the semi-classical spectrum (\ref{sc}) to be a 
valid description
of the full quantum theory only for large
quantum numbers. To gain a better understanding of this, one can 
perform a different kind of semi-classical analysis motivated by
two dimensional large-$N_c$ gauge theory. The expression (\ref{ham}) is 
isomorphic to the lightcone Hamiltonian of two-dimensional large-$N_c$
gauge theory minimally coupled to fundamental and adjoint matter particles 
at the endpoints
and interior points respectively. The origin of the linear string potential is
then the Coulomb force in two dimensions. Introducing the 
lightcone wavefunction
$\phi_N(x_0, x_1, \cdots, x_{N+1})$ for the sector with $N$ interior points,
with $x_j P^+ = P^{+}_{j}$,  
this linear potential between bosonic particles $j$ and $j+1$ becomes 
\cite{numerical}
\begin{eqnarray}
&& \frac{\sigma}{4\pi P^+} 
\int_{0}^{x_j+x_{j+1}} \  \frac{(x_j +y)(x_j+2x_{j+1}-y)}{
(x_j-y)^2 \sqrt{y x_j x_{j+1} (x_j+x_{j+1} -y)}}  
\nonumber \\
&& \left\{\phi_N(x_0, \cdots, x_{N+1}) \right. \nonumber \\
&& - \left. 
\phi_N(\cdots, x_{j-1}, y, x_j + x_{j+1} -y,\cdots)\right\} \ dy \nonumber \\
&& + \left[\frac{\sigma}{4P^+ \sqrt{x_j x_{j+1}}}\right]
\phi_N(x_0, \cdots, x_{N+1})
\end{eqnarray}
(In the gauge theory there would also be particle number changing interactions
that are not present in the first quantized string theory). 
The highly excited spectrum consists of wavefunctions $\phi_N$ that
oscillate rapidly and, following 't Hooft \cite{hoof2} and Kutasov 
\cite{kutasov}, one may 
simplify the integrand above in this limit. The integrals average to
zero except near the singularities of the integrand $x_j \approx y$,
in which case they may be effectively replaced by
\begin{eqnarray}
&& \frac{\sigma}{\pi P^+} 
\int_{-\infty}^{\infty} \frac{dz}{z^2} \  \left\{ 
\phi_N(x_0, \cdots, x_{N+1}) \right. \nonumber \\
&& \left. - \phi_N(\cdots, x_{j-1}, x_j + z, x_{j+1}-z,\cdots)\right\} \ . 
\end{eqnarray}
In the same regime, one can also assume that the mass terms are 
negligible compared to the 
excitation
energy. However, they do impose the boundary 
condition that $\phi_N = 0$ whenever any $x_j = 0$. 
Thus, in the $N=0$ sector the spectral 
eigenvalue equation for highly
excited states becomes
\be
M^2 \phi_0 = 
 \frac{2\sigma}{\pi} 
\int_{-\infty}^{\infty} dz \  \frac{\phi_0(x_0, x_{1}) - 
\phi_N(x_0 + z, x_1-z)}{z^2} \ . 
\eq
The solutions are of the form  
\begin{eqnarray}
\phi_0(x_0,x_1) & = & 
\sin s \pi x_0 \\
x_1 & = & 1-x_0 \\
M^2 & = & 2 \pi \sigma s
\end{eqnarray}
for large {\em positive} integers $s$. Comparing with eq(\ref{sc}), we
can identify $l_1 \equiv s$ in the $N=0$ sector. Note, however, that
the new analysis excludes the troublesome $l_1=0$ solution. 

Proceeding in the same way for $N=1$, the corresponding 
eigenvalue equation for highly
excited states becomes
\begin{eqnarray}
M^2 \phi_1 & = &
 \frac{2\sigma}{\pi} 
\int_{-\infty}^{\infty} \frac{dz}{z^2} \  
\left\{ \phi_1(x_0, x_{1}, x_2) \right. \nonumber \\
&&  - \phi_1(x_0 + z, x_1-z,x_2)  \nonumber \\
& &   +
\phi_1(x_0, x_{1}, x_2) \nonumber \\ 
&& - 
\left. \phi_1(x_0, x_1+z,x_2-z)\right\}  \ . 
\end{eqnarray}
The solutions in this case which respect the boundary conditions are
\begin{eqnarray}
\phi_1(x_0,x_1,x_2) & = & 
\sin s_1 \pi x_0 \sin s_2 \pi x_2 \nonumber \\
&& \pm \sin s_1 \pi x_2 \sin s_2 \pi x_0  
\\
x_0 + x_1 + x_2  & = & 1 \\
M^2 & = & 2 \pi \sigma (s_1 + s_2)
\end{eqnarray}
for large positive integers $s_1$ and $s_2$ such that $s_2>s_1$.
The plus solution occurs when $s_2-s_1$ is odd,
the minus solution when it is even, corresponding to states of opposite
parity under orientation reversal of the open string. The identification
with eq(\ref{sc}) is made by $l_2 = s_1$, $l_1 + l_2 = s_2$. Again, the
solutions $l_1=0$ and $l_2=0$ are excluded, but this time one can use large
$s_1$ and $s_2$, where the analysis is valid, to exclude the former.

\subsection{Asymptotic Density of States}

The natural generalisation of these results is that in the $N$-sector
the excited spectrum is
\be
M^2 =  2 \pi \sigma (s_1 + s_2 + \cdots + s_{N+1})
\label{Nspec}
\eq
for large positive integers $s_{N+1} > s_{N} > \cdots > s_1$, with
\be
s_j = \sum_{n=N-j+2}^{N+1} l_n \ .
\eq
This matches (\ref{sc}) provided $l_n = 0$ is excluded.
Note that a similar asymptotic spectrum has been derived 
for closed strings of bosonic and fermionic adjoint matter in 
refs.\cite{kutasov,bvds}, although only 
the solutions even under orientation reversal were found.
Numerical solutions for the low-lying spectrum 
of two dimensional large-$N_c$ gauge theory coupled
to adjoint matter have also been obtained \cite{numerical}.  
They provide additional support for the argument that each $N$-sector 
should contribute
independently to the spectrum, even in the $\mu \to 0$ limit,
since no  degeneracies across
different $N$-sectors are observed.

The asymptotic density of states corresponding to the spectrum (\ref{Nspec}),
including all $N$-sectors, can be obtained in a standard way from the 
generating function
\be
G(w) \equiv \sum_{n=1}^{\infty} d_n w^n = \prod_{m=1}^{\infty} (1+w^m)  
\eq
where $d_n$ is the number of states at level  $M^2 = 2 \pi \sigma n$.
The large $n$ behaviour is obtained from the limit $w \to 1$:
\begin{eqnarray}
\ln G & = & - \sum_{m,q=1}^{\infty} \frac{(-w^m)^q}{q} \nonumber \\
& = & -\sum_{q=1}^{\infty} \frac{(-w)^q}{q(1-w^q)} \nonumber \\
& \to  & \frac{1}{1-w} \sum_{q=1}^{\infty} \frac{(-w)^q}{q^2} \ \ ({\rm as} \
w \to 1) \nonumber \\
& = & \frac{\pi^2}{12(1-w)} \ .
\end{eqnarray}
Then $d_n$ can be obtained from the saddle point approximation to the
integral
\be
d_n = \frac{1}{2 \pi {\rm i}} \int_{\cal C} \frac{G(w)}{w^{n+1}}
\eq
where the contour ${\cal C}$ encircles the origin. The result
\be
d_n \sim {\rm exp}\left[ M \sqrt{\frac{\pi}{6 \sigma}} \right]
\eq
is characteristic of a $c=1/2$ conformal field theory (eq(\ref{exp})).

In fact, one could have guessed this result without further calculation
by rewriting the spectrum (\ref{Nspec}), including all $N$-sectors, as
\be
M^2 = 2 \pi \sigma \sum_{n=1}^{\infty} n l
\eq
where $l \in \{0,1\}$. This is the spectrum of a tower of fermionic
harmonic oscillators! The non-linear massive bosonic longitudinal oscillations 
of this string model contribute, in the asymptotic spectrum, just as
would a single free massless Majorana worldsheet fermion field. 
This is the right
number of degrees of freedom to account for the lowering of the Hagedorn
temperature observed in large-$N_c$ lattice simulations.

\section{Discussion}

Lattice data that may be used
to constrain any purported string theory of QCD have been critically reviewed. 
In general, one must extrapolate data to $N_c = \infty$ before comparing with 
free-string formulas. From data on the Luscher coefficient 
and Hagedorn temperature, in additional to the usual massless
degrees of freedom associated with oscillations in $D-2$ transverse 
dimensions, it appears that 
further massive modes contribute to the asymptotic spectrum.
Subject to the numerical accuracy of the data, they contribute
effectively like a $c \approx 1/2$ conformal worldsheet field. 

One obvious place to look for the extra modes is in the longitudinal
direction. The asymptotic spectrum of an `old' 
massive relativistic string model in two dimensions was re-derived and 
clarified.
It possesses only longitudinal oscillations. Despite being a 
bosonic model, the counting of highly excited states matches 
that of a $c=1/2$ worldsheet conformal field. 
The observations made in this paper are also relevant for `new' approaches
to string theory. Longitudinal modes of oscillation will exist in the
bulk for holograhic models based on the ADS/CFT correspondence
\cite{brower}. The constraints
from lattice data may help to identify the correct background that
corresponds to confining gauge theory. In this and
other string models, there still 
remains much to understand concerning the usual unitarity, space-time
symmetry, and groundstate stability expected of gauge theory. 

To test the hypothesis suggested in this paper, it would be useful to 
have further lattice data at large-$N_c$ for
more detailed observables. For example, the thermodynamic pressure 
may vary rapidly-enough close to the Hagedorn temperature for a quantitative
comparison to be made with the result based on the density of states
(\ref{exp}) with $c = (D-2) + 1/2$ and $D_\perp = 2$. In order to make 
general statements, in this paper we have
essentially been considering data from the large-$l$ and minimum-$l$ 
behaviour of the groundstate functions $E_{c}(l)$, $E_o(l)$ 
in eqs.~(\ref{open})(\ref{closed}). 
Of course,
it would be interesting to compare the 
predictions of specific string models and
large-$N_c$ gauge theory at general $l$ and for low excited states of these
systems \cite{morning}.

\acknowledgments{
The work is supported by PPARC grant 
PP/D507407/1. I thank B. Lucini for helpful discussions.  }

\end{document}